\documentclass[a4paper,11pt]{article}
\usepackage{pos}
\usepackage{upgreek}
\usepackage{bm}
\usepackage{subfigure} 
\usepackage{hyperref}
\usepackage{wrapfig}
\usepackage{setspace}
\usepackage{enumitem}

\title{Search for UHE neutrinos from GRBs with the Pierre Auger Observatory}
 \ShortTitle{Search for UHE neutrinos from GRBs with the Pierre Auger Observatory}

\author*[ab]{Yago Lema-Capeans}

\affiliation[a]{Instituto Galego de Física de Altas Enerxías (IGFAE), Universidade de Santiago de Compostela,
Santiago de Compostela, Spain}

\onbehalf{for the Pierre Auger Collaboration$^b$}
\affiliation[b]{Observatorio Pierre Auger, Av.\ San Mart{\'\i}n Norte 304, 5613 Malarg\"ue, Argentina\\
Full author list: {\rm\url{https://www.auger.org/archive/authors_icrc_2025.html}}}

\emailAdd{spokespersons@auger.org}

\abstract{We report on the search for ultra-high-energy neutrinos from the prompt emission of gamma-ray bursts (GRBs) using Surface Detector (SD) data from Phase I of the Pierre Auger Observatory (2004–2021). A total of 570 GRBs occur within the most neutrino-sensitive field of view of the SD, considering both Earth-skimming and downward-going detection channels. For this purpose, GRB neutrino emission has been modeled using the numerical software NeuCosmA, incorporating gamma-ray measurements and inferred parameters such as the jet Lorentz factor and the minimum variability time scale. No neutrino candidates were found, and upper limits were obtained by stacking the individual GRB neutrino fluences. These limits are complementary to those of IceCube and ANTARES and provide the strongest constraints on prompt GRB neutrino fluence above  $10^{18}$ eV. Additionally, limits on GRB fluence in alternative models of neutrino production have been derived using Auger data.}

\ConferenceLogo{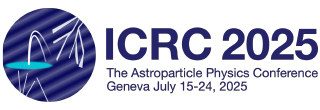}

\FullConference{%
39th International Cosmic Ray Conference (ICRC2025)\\
15 -- 24 July 2025\\
Geneva, Switzerland}

\begin{document}
\maketitle

\section{Introduction}

 The sources of Ultra-High Energy (UHE) cosmic rays, although confirmed to be extragalactic \cite{dipole}, remain unknown. Among them, Gamma-Ray Bursts (GRBs) are promising candidates due to their extreme energy release over time scales of seconds to minutes. 
 Proposals to explain the gamma-ray emission from GRBs also predict UHE neutrinos as the result of collisions of accelerated protons and nuclei with matter and photons in the GRB \cite{WaxBah}. Their detection would be a major step toward identifying UHE cosmic-ray sources and Multimessenger Astronomy. Since no neutrino candidates were observed at the Pierre Auger Observatory between 1 Jan 2004 and 31 Dec 2021, we place upper limits on the UHE neutrino fluence from GRBs.

\section{Models of prompt neutrino emission from GRBs}

\begin{wrapfigure}{r}{0.50\textwidth}
  \centering
  \includegraphics[width=0.50\textwidth]{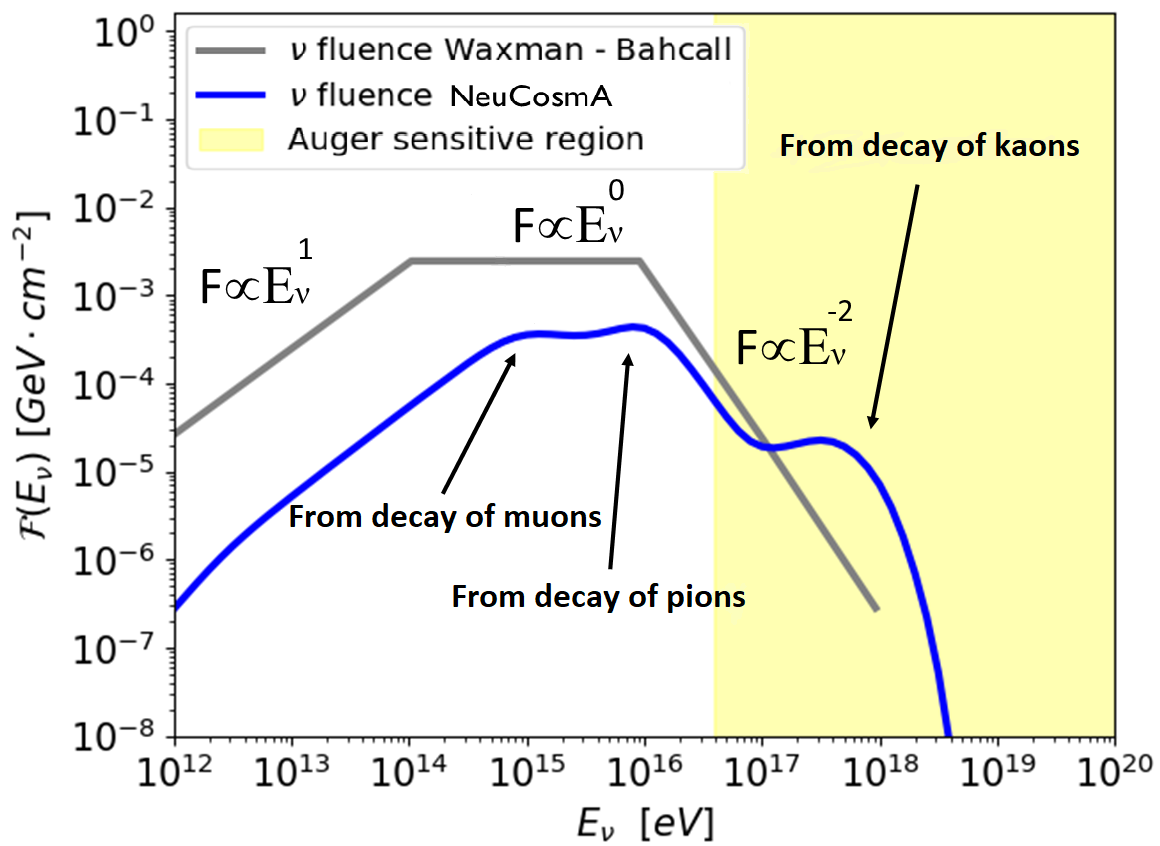}
  \caption{The predicted Waxman-Bahcall (WB) neutrino fluence for an individual GRB \cite{WaxBah}, compared to the fluence obtained for an average GRB with the NeuCosmA numerical code (blue line). The yellow region indicates the energy range where the Pierre Auger Observatory is sensitive to showers induced by neutrinos.}
  \label{fig:WBvsNeuCosmA}
\end{wrapfigure}

One of the most popular models for the gamma-ray emission in GRBs \cite{Meszaros} is the fireball model. In this model, the thermal pressure arising from the central engine drives an expanding fireball through the surrounding material. Differences between velocities of shells of material ejected at different times creates shocked regions where Fermi-accelerated electrons emit photons through synchrotron radiation, yielding the bulk of the gamma-ray emission in GRBs. This is the so-called prompt photon emission.

Neutrinos are thought to be generated by protons or nuclei, which are also expected to be Fermi-accelerated in the internal shocks after collisions with the radiation and matter in the GRB jet. In the Waxman-Bahcall (WB) pioneering model of neutrino production in GRB \cite{WaxBah}, neutrinos would be created mainly by the decay of charged pions and muons by photoproduction processes via the $\Delta^{+}$ resonance $\mathrm{p} + \upgamma \rightarrow \Delta^{+} \rightarrow \mathrm{n} + \uppi^{+}$ for $1/3$ of all cases.

The present work is mainly focused on testing the internal shock model (IS) with UHE neutrinos beyond the WB analytical picture. For this purpose, we first adopted the numerical estimation of neutrino fluxes of the `Neutrinos from Cosmic ray Accelerators' (NeuCosmA) software \cite{Fireball_revisited, Models_NeuCosmA}. Using this software, besides neutrinos from the decay of the charged pions
through the $\Delta^{+}$ resonance, 
contributions from higher-energy resonances, direct $\mathrm{t}$-channel and multipion production are also considered. Moreover, not only neutrinos from the decay of pions and muons are accounted for, but also those produced in the decay of neutrons and kaons are included, 
leading to a neutrino energy spectrum that differs from the original WB prediction, as depicted in Fig.\,\ref{fig:WBvsNeuCosmA}. The spectrum exhibits a first peak at low energy related to the neutrinos produced in muon decays, a second one related to the decay of charged pions and a peak at high energy above $10^{17}\,$eV that can be tracked to the decay of kaons, which leads to a higher fluence at energies relevant to the Pierre Auger Observatory. The adopted version of NeuCosmA simulates GRBs with average shell properties assuming a representative one-zone collision occurring at the internal shock radius ($R_\mathrm{IS} \sim 10^{12-13}\, $cm) .

In the NeuCosmA fireball framework, we only consider pure primary proton spectra with a power-law of $E^{-2}$. However, heavier nuclei may also be synthesized in GRBs, potentially leading to different predictions for high- and ultra-high-energy neutrino emission~\cite{Lia}. 
In this work, we additionally consider alternative GRB emission scenarios that include nuclear cascades, specifically, the photospheric model, the IS (as described above, but including nuclei), and the internal-collision-induced magnetic reconnection and turbulence (ICMART) model. A key difference among these models lies in the characteristic radii at which radiation is efficiently emitted, which in turn determines the photon number density and, hence, the efficiency of neutrino production.
In the photospheric model, the bulk of the radiation is produced in the optically thick region below the photosphere ($R_\mathrm{ph} \sim 10^{11\text{--}12}\,$cm), where dissipative processes shape the photon spectrum. The ICMART model, on the other hand, assumes a Poynting-flux-dominated jet composed of magnetized shells that undergo magnetic reconnection upon collision. In this scenario, energy dissipation occurs at much larger distances ($R_\mathrm{ICMART} \sim 10^{15}\,$ cm$\,$\hspace{-0.05cm}).

\vspace{-0.16cm}
\section{Neutrino fluence calculation 
and GRB sample}
\vspace{-0.16cm}
NeuCosmA normalizes the neutrino fluence \textit{for each GRB}, using the observed $\upgamma$-ray fluence, and requires several input parameters. Firstly, the data obtained from the $\upgamma$-ray measurements of satellites, namely the duration of the GRB given by $T_{90}$ and the $\upgamma$-ray fluence in the corresponding energy band of the detector. It also uses information about the model that best fits the $\upgamma$-ray spectral data, in particular the power-law indices (low energy $\alpha$, high energy $\beta$) and the energy break ($E_p$). 

In addition to this set of parameters, the internal shock radius $R_\mathrm{IS}$ and the normalization of the neutrino fluence obtained through NeuCosmA depend on other relevant astrophysical parameters that are either measured only for a small fraction of GRBs, or are unknown, such as the redshift, the minimum variability time scale, and the Lorentz factor of the GRB jet.

\begin{figure}[ht]
\centering
\subfigure{\includegraphics[width=0.495\linewidth]{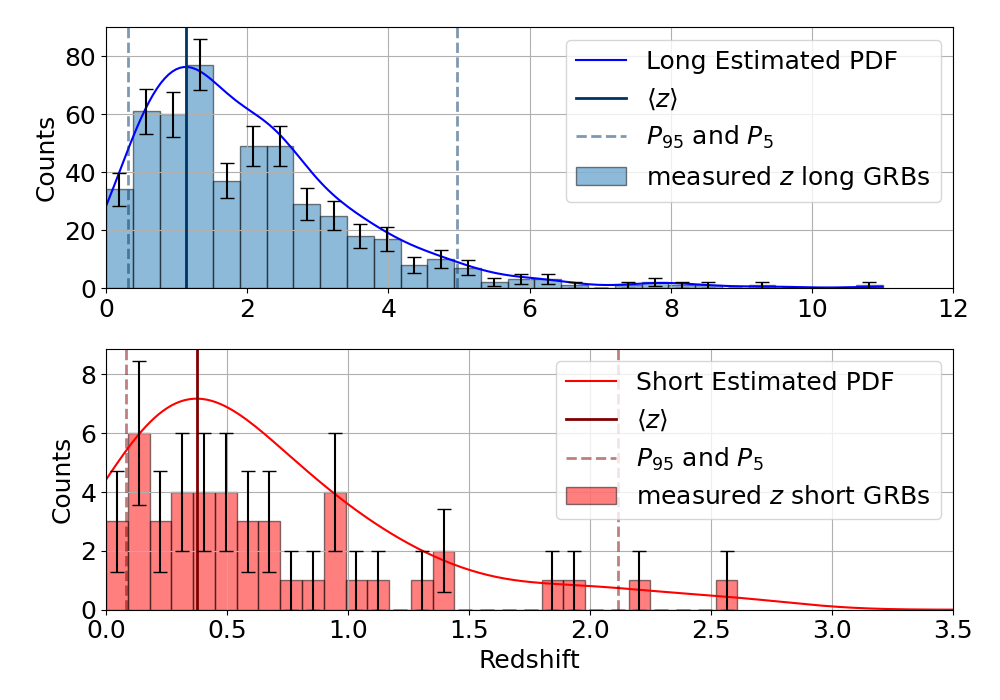}}
\subfigure{\includegraphics[width=0.49\linewidth]{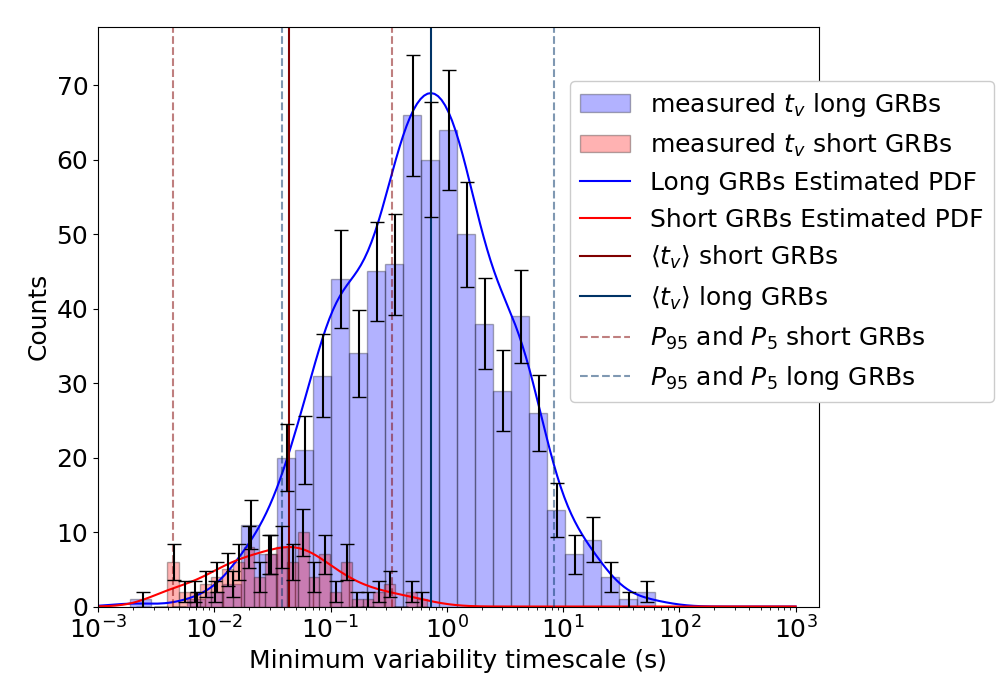}}
\caption{Redshift $z$ (left) and minimum variability timescale $t_v$ (right) distributions for all GRBs with available data. Red for short GRBs and blue for long GRBs. The thick lines correspond to the estimated PDFs (see text).
}
\label{fig:PDFs}
\end{figure}

The \textbf{redshift}, $z$, is usually inferred from GRB \textit{afterglow} measurements and is a main source of uncertainty in the estimation of GRB neutrino emission. In the GRB sample used (570 GRBs, see below), $z$ is known for a fraction of 16\%. Due to the limited availability of $z$ measurements, probability density functions (PDFs) for $z$ are separately estimated for long and short GRBs based on the redshift distributions of those with known values. A Gaussian kernel density estimation method is used to infer the PDFs from the distributions. As shown in the left panel of Fig.\,\ref{fig:PDFs}, short GRBs are typically observed closer to Earth, exhibiting an average redshift $\langle z_s \rangle \simeq 0.67$, while for long GRBs $\langle z_l \rangle \simeq 2.03$. We assume that the inferred PDFs in $z$ are representative of the whole population of observed long and short GRBs, respectively. We note that the subset of GRBs with measured redshift may be biased towards lower redshifts since afterglow observations -and hence spectroscopic follow-up- become more challenging for high-$z$ events. These distributions are used afterwards to obtain, for each GRB, the average GRB neutrino fluence and its uncertainty band.

The \textbf{minimum variability timescale} $t_v$ of the GRB photon lightcurve is directly related to the size of the source and determines the p$\gamma$ optical depth. It can be inferred using Haar wavelets on the temporal properties of the lightcurves of each GRB 
\cite{MVT_2}. This parameter is also an important source of uncertainty and is only known for about 12\% of GRBs in the sample. We use the same methodology applied for the redshift, constructing PDFs for both long and short GRBs from the distributions of measured values of $t_v$. The resulting PDFs are shown in the right panel of Fig.\,\ref{fig:PDFs}. 
The average values of $t_v$ are different for the two populations of GRBs, $\langle t_v\rangle \sim 37\,\mathrm{ms}$ for short GRBs and $\langle t_v\rangle \sim 0.59\,\mathrm{s}$ for long GRBs.

The \textbf{bulk Lorentz factor} $\Gamma$ of the GRB expanding shells (of the GRB jet) 
can be estimated in very bright GRBs through the detection of an early peak in the afterglow optical or in the GeV lightcurve \cite{Ghirlanda}. For these GRBs, $\Gamma$ represents the jet Lorentz factor of the ejecta before this transition. Only 6\% of the GRBs in the sample have an estimated value of $\Gamma$. Instead of building yet another PDF, we have estimated $\Gamma$ for each GRB, taking advantage of an existing empirical correlation \cite{Ghirlanda} between $\Gamma$ and the bolometric ([1 keV, 10 MeV]) isotropic equivalent $\gamma$-ray luminosity $L_\mathrm{iso}$  
\begin{equation}
    \Gamma = 182  \left(\frac{L_\mathrm{iso}}{10^{52} \,\mathrm{erg} \cdot \mathrm{s}^{-1}}\right)^{0.324}.
    \label{eq:empirical}
\end{equation}
We estimate $L_{\mathrm{iso}}$ using a method that accounts for k-corrections, integrating the GRB spectral model available from databases in the bolometric band. For Swift GRBs, for which only the lower energy part of the spectrum is measured due to its narrow energy range, the energy break $E_p$ was estimated using another empirical correlation between $E_p$ and the power-law index $\alpha_\mathrm{BAT}$ \cite{EpAlpha}.

The fraction of energy carried by the magnetic field with respect to the energy in electrons, $E_B/E_e$, affecting the synchrotron loss time, is unknown, and no method exists so far to probe it. In this work, we have assumed a log-uniform PDF around equipartition with values from 0.1 to 10. The same happens with the baryonic loading factor, $f_p$, representing the energy in the fireball going into protons with respect to that into electrons. We fix it to a reference value of $f_p = 10$ \cite{Fireball_revisited}.

For each GRB in the sample, an average neutrino fluence is obtained directly with NeuCosmA using either measured parameters or the means of the corresponding distributions. Then, an uncertainty band is obtained by simulating 1000 fluences per GRB, sampling the parameters either within the uncertainties of the measurements if known or from the corresponding PDFs in case they are unknown. The parameters of the two empirical relations considered in this work ($\Gamma-L_{\mathrm{iso}}$) and ($E_p-\alpha_\mathrm{BAT}$) are also sampled through PDFs built from likelihoods of the corresponding fits to propagate the corresponding uncertainty. The fluence uncertainty band is constructed using the 5\% and 95\% percentiles in each neutrino energy bin, thereby containing 90\% of the simulated fluences. 

For the case of the Photospheric, IS and ICMART models with nuclear cascade, a simulation is not performed as was done with NeuCosmA in the IS with protons only. Instead, 
and for reference, we use the neutrino fluences shown in Fig.6 in \cite{Lia} corresponding to GRB jets loaded with ${}^{56}$Fe with an $E^{-2}$ energy spectrum for the parent nuclei,  
for \textit{fixed} values of $\Gamma = 300$, $t_v = 0.5\,\mathrm{s}$, time duration $t_\mathrm{dur}=30\,\mathrm{s}$ and $z=2$. We then rescale the fluence on Earth to account for the different $z$ and  $t_\mathrm{dur}=\mathrm{T_{90}}$ (see eq.(4.14) in \cite{Lia}), so that the number of particles produced per unit volume, time and per energy interval in the jet rest frame is the same for all GRBs in the sample. Photospheric and IS models also have nuclei loading factors $f_A = 10$, accounting for the fraction of energy in accelerated nuclei. The ICMART model is expected to be more efficient in accelerating particles and to have $f_A = 1$ from higher values of efficiencies in electron and proton acceleration (see Table 1 in \cite{Lia}).

The primary GRB dataset for this study was collected from the IceCube tool GRBweb\footnote{GRBweb available at this link \href{https://user-web.icecube.wisc.edu/~grbweb_public/}{GRBweb}. FERMIGBRST catalog available at this link \href{https://heasarc.gsfc.nasa.gov/db-perl/W3Browse/w3table.pl?tablehead=name\%3Dfermigbrst&Action=More+Options}{Fermi}. Swift catalog available at this link \href{https://swift.gsfc.nasa.gov/archive/grb_table}{Swift}. GCN Circulars available at this link \href{https://gcn.nasa.gov/circulars}{GCN}.}, providing a catalog of GRBs combining information from databases from various detectors. 
A dataset of 4369 GRBs was initially obtained from which direction, time of detection, $\upgamma$-ray fluence, $T_{90}$ and $z$ were extracted when available. 
Additionally, databases of the Fermi and Swift satellites were used to extract the parameters of the best fit to the measured $\upgamma$-ray spectrum. 
For those GRBs with no available spectral data, Konus-Wind information was extracted from GCN Circulars. 

From this initial sample, we discarded GRBs outside the SD Field of View (FoV), namely, with zenith angles as viewed from the SD $\theta\in[75^\circ,\,95^\circ]$, as well as those occurring during dead-time periods of the SD array. 
We further required each GRB to have a measured $\upgamma$-ray spectrum with an available best-fit model that provides the corresponding spectral parameters. We also discarded GRBs with very high uncertainty in the lower energy power-law index requiring $S(\alpha) < 1$. After these cuts, the sample comprises 140 GRBs in the Earth-Skimming (ES) channel with $\theta\in[90^\circ,\,95^\circ]$ and 430 GRBs in the downward-going (DGH) channel $\theta\in[75^\circ,\,90^\circ]$.

\section{Stacking procedure: calculation of limits}

The effective area of the Auger SD is obtained by integrating over neutrino energy $E_\nu$, the SD aperture multiplied by the neutrino cross section for each neutrino flavour ($i=\mathrm{e}, \upmu, \uptau$) and interaction channel (Charged-Current (CC) \& Neutral-Current (NC)),  weighted by the neutrino selection and detection efficiency obtained from Monte Carlo simulations \cite{Pointlike}. 
To calculate the exposure $\mathcal{E}_{\mathrm{GRB}}$ to a specific GRB, we integrate the effective area in a time window $T_{90}$ starting at the detection time, taking into account the variation in zenith angle $\theta(t)$ of the GRB. The expected number of neutrino events per GRB with a flux $\phi_{\mathrm{GRB}, i}(E_\nu)$ and interaction type $c$ = (CC, NC), is
\begin{equation}
    \mathcal{N}_{\mathrm{GRB}, i, c} =  \int_{E_{\min }}^{E_{\max }} \phi_{\mathrm{GRB}, i}(E_\nu)\, \mathcal{E}_{\mathrm{GRB}, i,c} (E_{\nu})  \mathrm{~d}E_{\nu}.
    \label{eq:Number_expected_nu}
\end{equation}
In the case of a GRB occurring in the FoV of the DGH channel, the total number of expected events is the sum of the individual $\mathcal{N}_{\mathrm{GRB}, i, c}$ for all flavours and interaction channels, whereas for a GRB in the ES channel, only the contribution from $\uptau-$CC is relevant. We denote the neutrino flux as $\phi_{\mathrm{GRB}, i}(E_\nu) = \phi_0 f_{\mathrm{GRB}, i}(E_{\nu})$, where $f_{\mathrm{GRB}, i}(E_{\nu})$ is a dimensionless function that encodes the energy dependence of the flux and is different for each GRB and flavour as predicted by NeuCosmA, and $\phi_0$ is a normalization factor. The expected number of neutrino events $\mathcal{N}_{\mathrm{GRB}}$ for the whole sample is the sum of all GRBs in the ES and DGH channels. Taking the Feldman-Cousins factor of 2.44 for an observation of 0 neutrinos with 0 background assumed, we can constrain $\phi_0$ at 90\% C.L.   
\begin{equation}
    \phi_0  = \frac{2.44}{{\displaystyle \sum_{{\mathrm{DGH \, GRB}}}\int_{E_{\min }}^{E_{\max }} \, f_{\mathrm{GRB}}(E_\nu)\, \mathcal{E}_{\mathrm{GRB}}^{\mathrm{DGH}} (E_{\nu})  \mathrm{~d}E_{\nu}  + \sum_{{\mathrm{ES \, GRB}}}\int_{E_{\min }}^{E_{\max }} \,  f_{\mathrm{GRB}, \uptau}(E_\nu)\, \mathcal{E}_{\mathrm{GRB}}^{\mathrm{ES}} (E_{\nu})  \mathrm{~d}E_{\nu}}}\, ,
\end{equation}
where $f_{\mathrm{GRB}}(E_\nu)\, \mathcal{E}_{\mathrm{GRB}}^{\mathrm{DGH}} (E_{\nu}) = \sum_{i,c}f_{\mathrm{GRB}, i}(E_{\nu}) \, \mathcal{E}_{\mathrm{GRB}, i, c}^{\mathrm{DGH}} (E_{\nu})$. 

The fluence for a single GRB $\mathcal{F}_{\mathrm{GRB}, i}$, proportional to the energy per unit area, is obtained  multiplying the flux $\phi_{\mathrm{GRB}, i}$ by $T_{90}$ and $E_{\nu}^2$. We compute a stacking limit on the fluence for the whole sample by adding the contributions of all GRBs normalized with $\phi_0$. The results can be shown in terms of fluence per GRB,  dividing the total fluence by the number of GRBs, $N_{\rm GRB}$,
\begin{equation}
    \mathcal{F}_{\mathrm{stack}} (E_{\nu})= 
    \frac{1}{N_{\rm GRB}} \sum_{\mathrm{GRB}, i}  \mathcal{F}_{\mathrm{GRB}, i} = 
    \frac{E^2_{\nu}}{N_{\rm GRB}} \sum_{\mathrm{GRB}, i} \phi_0 \cdot f_{\mathrm{GRB},i}(E_\nu) \,  T_{90, {\rm GRB}} \,, 
    \label{eq:Stack_fluence}
\end{equation}
where $f_{\mathrm{GRB}}(E_\nu) = \sum_i f_{\mathrm{GRB}, i}(E_\nu)$ and, by construction, the dependence on neutrino energy of the fluence limit follows the shape of the average fluence of GRBs in the sample. It is also possible to convert the fluence $\mathcal{F}_{\mathrm{stack}}$ into a \textit{quasi-diffuse} neutrino flux by multiplying it by the average rate of GRBs expected per year over the full sky 
\begin{equation}
\Phi(E_\nu) = \mathcal{F}_{\mathrm{stack}} (E_\nu)\cdot \frac{1}{4\pi} \cdot 667 \,\text{yr}^{-1}\,.
\label{eq:Stack_flux}
\end{equation}
\vspace{-1cm}
\section{Results and discussion}

\begin{figure}[ht]
\centering
\includegraphics[width=1  \linewidth]{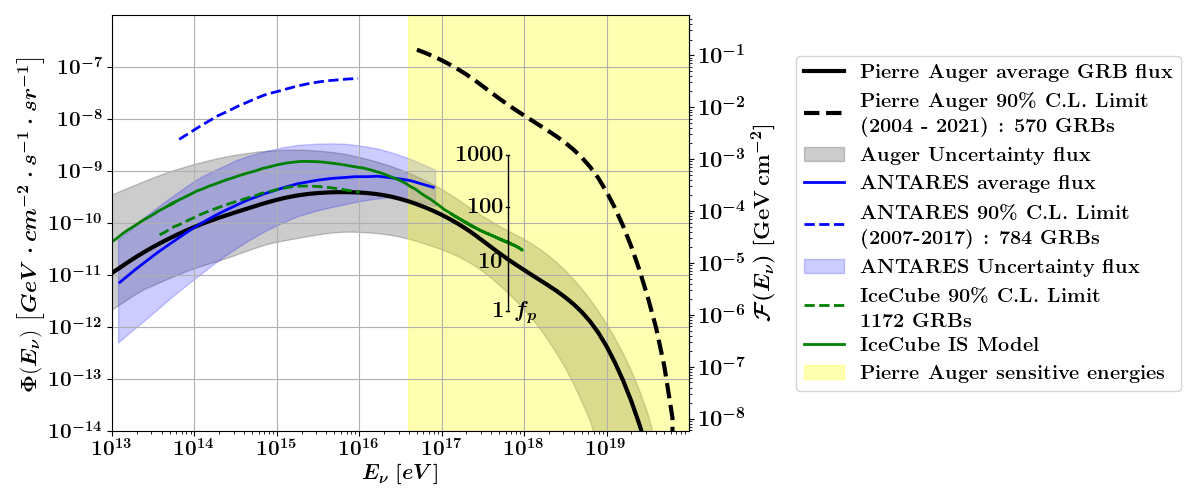}

\caption{Solid lines: Quasi-diffuse all-flavor neutrino flux (left-hand axis) and average fluence per GRB (right-hand axis) expected from the prompt phase of GRBs obtained with NeuCosmA for the sample of GRBs in the FoV of the Pierre Auger Observatory (this work, black line) and ANTARES \cite{Antares_Zegarelli} (blue). Dashed lines represent the corresponding stacking limits $\Phi$ in Eq.\,\eqref{eq:Stack_flux} and $\mathcal{F}_{\rm stack}$ in Eq.\,\eqref{eq:Stack_fluence} at $90\%$ CL. 
The vertical scale shown inside the plot indicates the baryonic loading $f_p$ with $f_p=10$ assumed by default.
In green, the expected neutrino emission (solid) and corresponding stacking limit (dashes) reported by IceCube for an IS Model \cite{IceCube}, assuming average values of $\Gamma=300$ and $f_p=10$ for all GRB in the sample.}
\label{fig:Flux_Limits}
\end{figure}

Figure\,\ref{fig:Flux_Limits} shows the expected neutrino flux and average fluence in the prompt GRB phase, as obtained with NeuCosmA, along with the stacking limits on both the GRB neutrino flux (left axis) and average fluence per GRB (right axis).
The uncertainty band in the expected fluence is obtained by averaging the upper and lower bounds of all individual GRB bands. The ES channel alone contributes $\simeq 86\%$ to the Pierre Auger stacking limit. 
The vertical scale shown inside the plot indicates the baryonic loading $f_p$ with $f_p=10$ assumed by default. Changing the value of $f_p$ would shift the entire fluence prediction and uncertainty band up or down. The ratio of the stacking limit to the average fluence is $\simeq 1.8\cdot10^3$, implying that only high values of $f_p$ would be constrained by the Pierre Auger Observatory limits. 

For comparison, ANTARES \cite{Antares_Zegarelli}, also using a sample of GRBs throughout 10 years, and IceCube \cite{IceCube} using IS models are also shown in Fig.\,\ref{fig:Flux_Limits}. ANTARES stacking limits were also obtained using NeuCosmA to simulate individual fluences. In fact, when comparing the average fluence expected for the GRBs in the ANTARES FoV with the one in this work,  
the spectral trend seen is similar, and the fluences are compatible within uncertainties given the fact that the GRB sample is not the same in both observatories. The ANTARES and Pierre Auger stacking limits (dashed lines) are of comparable magnitude, constraining quasi-diffuse flux values $\Phi\sim 10^{-7}\,{\rm GeV\,cm^{-2}\, s^{-1} sr^{-1}}$ at $E_\nu\sim 10^{17}$ eV. On the other hand, the limits obtained by IceCube (green dashed line) are considerably more constraining in the same energy range covered by ANTARES. 
The Pierre Auger limit becomes the most constraining for UHE neutrinos around and above $\sim 10^{18}$ eV and is complementary to those of ANTARES and IceCube.  

\begin{figure}
  \centering
\includegraphics[width=0.81  \linewidth]{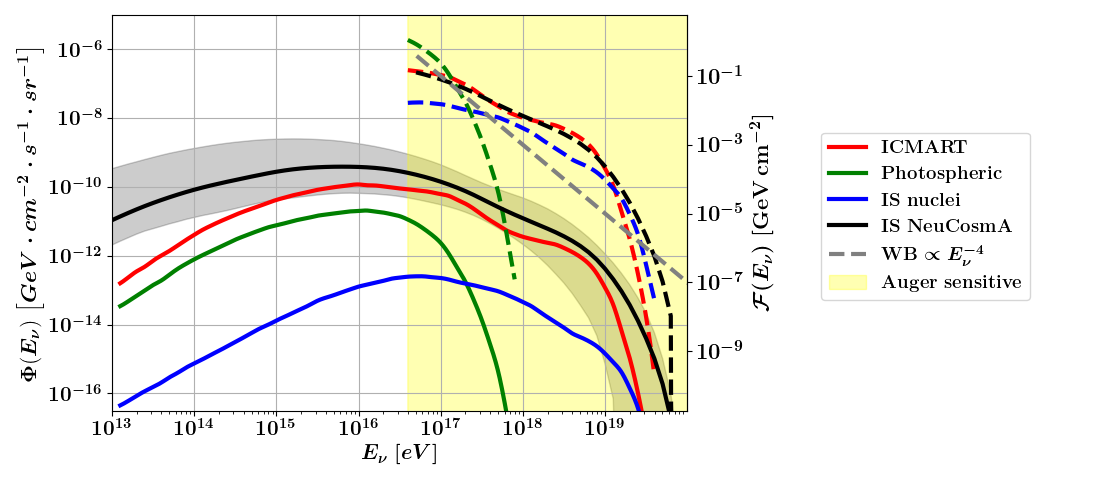}
\caption{Solid lines: Predicted quasi-diffuse all-flavor neutrino flux (left-hand axis) and average fluence per GRB (right-hand axis) for the photospheric (green), IS (blue) and ICMART (red) models with nuclear cascade, 
and $\Gamma = 300$, 
$z = 2$ and $f_A=10$ except for ICMART, $f_A=1$, and IS with NeuCosmA (no nuclear cascade, black). Dashed lines: corresponding stacking limits $\Phi$ in Eq.\,\eqref{eq:Stack_flux} and $\mathcal{F}_{\rm stack}$ in Eq.\,\eqref{eq:Stack_fluence} at $90\%$ CL. Limits to a flux $\propto E_\nu^{-4}$, as expected in the WB model, are also shown (gray).} 
\label{fig:Flux_Limits_Therese}
\end{figure}

Fig.\,\ref{fig:Flux_Limits_Therese} shows the diffuse neutrino flux and fluence predictions, along with the corresponding stacking limits for the Photospheric, IS and ICMART GRB models, as well as limits assuming a power-law neutrino spectrum $\propto E^{-4}$, as predicted by the analytical Waxman-Bahcall model at UHE. The normalization of the stacking limit is determined by the steepness in the energy of the neutrino flux predicted by the model. The fluence limits obtained are similar across all models, as they correspond to the level needed to yield 2.4 expected events in the detector. However, the ratio between these limits and the model predictions varies significantly, depending on the shape of each model's neutrino spectrum and the normalization of model prediction. This ratio reflects how strongly each model is constrained. The Photospheric model is the least constrained by the Pierre Auger Observatory, since its predicted neutrino flux falls sharply above $10^{18}$ eV, where the detector has maximum sensitivity. The new (IS) considering nuclear cascade is not very well constrained either because of the low normalization of the expected fluence. In contrast, the ICMART model is the most strongly constrained due to its relatively high predicted neutrino yield at ultra-high energies. The resulting ratio for this model is $\simeq 2.9\cdot10^{3}$, being slightly higher than that obtained for the Internal Shock (IS) model using NeuCosmA simulations. 

\newpage

In conclusion, stacking limits to the UHE neutrino flux were computed with NeuCosmA in the internal shock (IS) model, obtaining neutrino fluences for each GRB in the FoV of the Pierre Auger Observatory, based on satellite gamma-ray data. 
Expected average neutrino fluences and uncertainties were derived from 1000 simulations per GRB, with the jet Lorentz factor $\Gamma$ estimated via an empirical relation with luminosity, $z$ and $t_v$ values sampled from PDFs constructed from available measurements and fixed $f_p=10$.
The NeuCosmA-based stacking limits from the Pierre Auger Observatory are 
the most constraining at UHE, complementing those of other experiments.
Alternative GRB models including nuclear cascades, using benchmark fluences for standard parameters were also explored leading to similar conclusions.
\vspace{0.03cm}

{\bf Acknowledgments:} 
We gratefully acknowledge M. Bustamante and W. Winter for providing the NeuCosmA software used in this work.

\begingroup
\setstretch{0.12}

\endgroup

\clearpage

\section*{The Pierre Auger Collaboration}
{\footnotesize\setlength{\baselineskip}{10pt}
\noindent
\begin{wrapfigure}[11]{l}{0.12\linewidth}
\vspace{-4pt}
\includegraphics[width=0.98\linewidth]{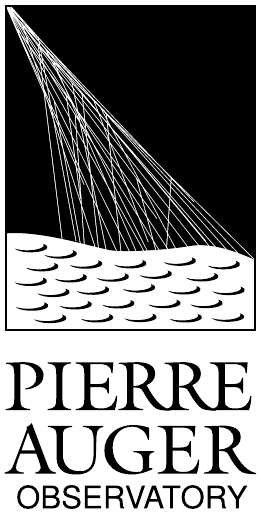}
\end{wrapfigure}
\begin{sloppypar}\noindent
A.~Abdul Halim$^{13}$,
P.~Abreu$^{70}$,
M.~Aglietta$^{53,51}$,
I.~Allekotte$^{1}$,
K.~Almeida Cheminant$^{78,77}$,
A.~Almela$^{7,12}$,
R.~Aloisio$^{44,45}$,
J.~Alvarez-Mu\~niz$^{76}$,
A.~Ambrosone$^{44}$,
J.~Ammerman Yebra$^{76}$
% ...

\end{sloppypar}
\begin{center}
\end{center}

\vspace{1ex}
\begin{description}[labelsep=0.2em,align=right,labelwidth=0.7em,labelindent=0em,leftmargin=2em,noitemsep,before={\renewcommand\makelabel[1]{##1 }}]
\item[$^{1}$] Centro At\'omico Bariloche and Instituto Balseiro (CNEA-UNCuyo-CONICET), San Carlos de Bariloche, Argentina
\item[$^{2}$] Departamento de F\'\i{}sica and Departamento de Ciencias de la Atm\'osfera y los Oc\'eanos, FCEyN, Universidad de Buenos Aires and CONICET, Buenos Aires, Argentina
\item[$^{3}$] IFLP, Universidad Nacional de La Plata and CONICET, La Plata, Argentina
\item[$^{4}$] Instituto de Astronom\'\i{}a y F\'\i{}sica del Espacio (IAFE, CONICET-UBA), Buenos Aires, Argentina
\item[$^{5}$] Instituto de F\'\i{}sica de Rosario (IFIR) -- CONICET/U.N.R.\ and Facultad de Ciencias Bioqu\'\i{}micas y Farmac\'euticas U.N.R., Rosario, Argentina
\item[$^{6}$] Instituto de Tecnolog\'\i{}as en Detecci\'on y Astropart\'\i{}culas (CNEA, CONICET, UNSAM), and Universidad Tecnol\'ogica Nacional -- Facultad Regional Mendoza (CONICET/CNEA), Mendoza, Argentina
\item[$^{7}$] Instituto de Tecnolog\'\i{}as en Detecci\'on y Astropart\'\i{}culas (CNEA, CONICET, UNSAM), Buenos Aires, Argentina
\item[$^{8}$] International Center of Advanced Studies and Instituto de Ciencias F\'\i{}sicas, ECyT-UNSAM and CONICET, Campus Miguelete -- San Mart\'\i{}n, Buenos Aires, Argentina
\item[$^{9}$] Laboratorio Atm\'osfera -- Departamento de Investigaciones en L\'aseres y sus Aplicaciones -- UNIDEF (CITEDEF-CONICET), Argentina
\item[$^{10}$] Observatorio Pierre Auger, Malarg\"ue, Argentina
\item[$^{11}$] Observatorio Pierre Auger and Comisi\'on Nacional de Energ\'\i{}a At\'omica, Malarg\"ue, Argentina
\item[$^{12}$] Universidad Tecnol\'ogica Nacional -- Facultad Regional Buenos Aires, Buenos Aires, Argentina
\item[$^{13}$] University of Adelaide, Adelaide, S.A., Australia
\item[$^{14}$] Universit\'e Libre de Bruxelles (ULB), Brussels, Belgium
\item[$^{15}$] Vrije Universiteit Brussels, Brussels, Belgium
\item[$^{16}$] Centro Brasileiro de Pesquisas Fisicas, Rio de Janeiro, RJ, Brazil
\item[$^{17}$] Centro Federal de Educa\c{c}\~ao Tecnol\'ogica Celso Suckow da Fonseca, Petropolis, Brazil
\item[$^{18}$] Instituto Federal de Educa\c{c}\~ao, Ci\^encia e Tecnologia do Rio de Janeiro (IFRJ), Brazil
\item[$^{19}$] Universidade de S\~ao Paulo, Escola de Engenharia de Lorena, Lorena, SP, Brazil
\item[$^{20}$] Universidade de S\~ao Paulo, Instituto de F\'\i{}sica de S\~ao Carlos, S\~ao Carlos, SP, Brazil
\item[$^{21}$] Universidade de S\~ao Paulo, Instituto de F\'\i{}sica, S\~ao Paulo, SP, Brazil
\item[$^{22}$] Universidade Estadual de Campinas (UNICAMP), IFGW, Campinas, SP, Brazil
\item[$^{23}$] Universidade Estadual de Feira de Santana, Feira de Santana, Brazil
\item[$^{24}$] Universidade Federal de Campina Grande, Centro de Ciencias e Tecnologia, Campina Grande, Brazil
\item[$^{25}$] Universidade Federal do ABC, Santo Andr\'e, SP, Brazil
\item[$^{26}$] Universidade Federal do Paran\'a, Setor Palotina, Palotina, Brazil
\item[$^{27}$] Universidade Federal do Rio de Janeiro, Instituto de F\'\i{}sica, Rio de Janeiro, RJ, Brazil
\item[$^{28}$] Universidad de Medell\'\i{}n, Medell\'\i{}n, Colombia
\item[$^{29}$] Universidad Industrial de Santander, Bucaramanga, Colombia
\item[$^{30}$] Charles University, Faculty of Mathematics and Physics, Institute of Particle and Nuclear Physics, Prague, Czech Republic
\item[$^{31}$] Institute of Physics of the Czech Academy of Sciences, Prague, Czech Republic
\item[$^{32}$] Palacky University, Olomouc, Czech Republic
\item[$^{33}$] CNRS/IN2P3, IJCLab, Universit\'e Paris-Saclay, Orsay, France
\item[$^{34}$] Laboratoire de Physique Nucl\'eaire et de Hautes Energies (LPNHE), Sorbonne Universit\'e, Universit\'e de Paris, CNRS-IN2P3, Paris, France
\item[$^{35}$] Univ.\ Grenoble Alpes, CNRS, Grenoble Institute of Engineering Univ.\ Grenoble Alpes, LPSC-IN2P3, 38000 Grenoble, France
\item[$^{36}$] Universit\'e Paris-Saclay, CNRS/IN2P3, IJCLab, Orsay, France
\item[$^{37}$] Bergische Universit\"at Wuppertal, Department of Physics, Wuppertal, Germany
\item[$^{38}$] Karlsruhe Institute of Technology (KIT), Institute for Experimental Particle Physics, Karlsruhe, Germany
\item[$^{39}$] Karlsruhe Institute of Technology (KIT), Institut f\"ur Prozessdatenverarbeitung und Elektronik, Karlsruhe, Germany
\item[$^{40}$] Karlsruhe Institute of Technology (KIT), Institute for Astroparticle Physics, Karlsruhe, Germany
\item[$^{41}$] RWTH Aachen University, III.\ Physikalisches Institut A, Aachen, Germany
\item[$^{42}$] Universit\"at Hamburg, II.\ Institut f\"ur Theoretische Physik, Hamburg, Germany
\item[$^{43}$] Universit\"at Siegen, Department Physik -- Experimentelle Teilchenphysik, Siegen, Germany
\item[$^{44}$] Gran Sasso Science Institute, L'Aquila, Italy
\item[$^{45}$] INFN Laboratori Nazionali del Gran Sasso, Assergi (L'Aquila), Italy
\item[$^{46}$] INFN, Sezione di Catania, Catania, Italy
\item[$^{47}$] INFN, Sezione di Lecce, Lecce, Italy
\item[$^{48}$] INFN, Sezione di Milano, Milano, Italy
\item[$^{49}$] INFN, Sezione di Napoli, Napoli, Italy
\item[$^{50}$] INFN, Sezione di Roma ``Tor Vergata'', Roma, Italy
\item[$^{51}$] INFN, Sezione di Torino, Torino, Italy
\item[$^{52}$] Istituto di Astrofisica Spaziale e Fisica Cosmica di Palermo (INAF), Palermo, Italy
\item[$^{53}$] Osservatorio Astrofisico di Torino (INAF), Torino, Italy
\item[$^{54}$] Politecnico di Milano, Dipartimento di Scienze e Tecnologie Aerospaziali , Milano, Italy
\item[$^{55}$] Universit\`a del Salento, Dipartimento di Matematica e Fisica ``E.\ De Giorgi'', Lecce, Italy
\item[$^{56}$] Universit\`a dell'Aquila, Dipartimento di Scienze Fisiche e Chimiche, L'Aquila, Italy
\item[$^{57}$] Universit\`a di Catania, Dipartimento di Fisica e Astronomia ``Ettore Majorana``, Catania, Italy
\item[$^{58}$] Universit\`a di Milano, Dipartimento di Fisica, Milano, Italy
\item[$^{59}$] Universit\`a di Napoli ``Federico II'', Dipartimento di Fisica ``Ettore Pancini'', Napoli, Italy
\item[$^{60}$] Universit\`a di Palermo, Dipartimento di Fisica e Chimica ''E.\ Segr\`e'', Palermo, Italy
\item[$^{61}$] Universit\`a di Roma ``Tor Vergata'', Dipartimento di Fisica, Roma, Italy
\item[$^{62}$] Universit\`a Torino, Dipartimento di Fisica, Torino, Italy
\item[$^{63}$] Benem\'erita Universidad Aut\'onoma de Puebla, Puebla, M\'exico
\item[$^{64}$] Unidad Profesional Interdisciplinaria en Ingenier\'\i{}a y Tecnolog\'\i{}as Avanzadas del Instituto Polit\'ecnico Nacional (UPIITA-IPN), M\'exico, D.F., M\'exico
\item[$^{65}$] Universidad Aut\'onoma de Chiapas, Tuxtla Guti\'errez, Chiapas, M\'exico
\item[$^{66}$] Universidad Michoacana de San Nicol\'as de Hidalgo, Morelia, Michoac\'an, M\'exico
\item[$^{67}$] Universidad Nacional Aut\'onoma de M\'exico, M\'exico, D.F., M\'exico
\item[$^{68}$] Institute of Nuclear Physics PAN, Krakow, Poland
\item[$^{69}$] University of \L{}\'od\'z, Faculty of High-Energy Astrophysics,\L{}\'od\'z, Poland
\item[$^{70}$] Laborat\'orio de Instrumenta\c{c}\~ao e F\'\i{}sica Experimental de Part\'\i{}culas -- LIP and Instituto Superior T\'ecnico -- IST, Universidade de Lisboa -- UL, Lisboa, Portugal
\item[$^{71}$] ``Horia Hulubei'' National Institute for Physics and Nuclear Engineering, Bucharest-Magurele, Romania
\item[$^{72}$] Institute of Space Science, Bucharest-Magurele, Romania
\item[$^{73}$] Center for Astrophysics and Cosmology (CAC), University of Nova Gorica, Nova Gorica, Slovenia
\item[$^{74}$] Experimental Particle Physics Department, J.\ Stefan Institute, Ljubljana, Slovenia
\item[$^{75}$] Universidad de Granada and C.A.F.P.E., Granada, Spain
\item[$^{76}$] Instituto Galego de F\'\i{}sica de Altas Enerx\'\i{}as (IGFAE), Universidade de Santiago de Compostela, Santiago de Compostela, Spain
\item[$^{77}$] IMAPP, Radboud University Nijmegen, Nijmegen, The Netherlands
\item[$^{78}$] Nationaal Instituut voor Kernfysica en Hoge Energie Fysica (NIKHEF), Science Park, Amsterdam, The Netherlands
\item[$^{79}$] Stichting Astronomisch Onderzoek in Nederland (ASTRON), Dwingeloo, The Netherlands
\item[$^{80}$] Universiteit van Amsterdam, Faculty of Science, Amsterdam, The Netherlands
\item[$^{81}$] Case Western Reserve University, Cleveland, OH, USA
\item[$^{82}$] Colorado School of Mines, Golden, CO, USA
\item[$^{83}$] Department of Physics and Astronomy, Lehman College, City University of New York, Bronx, NY, USA
\item[$^{84}$] Michigan Technological University, Houghton, MI, USA
\item[$^{85}$] New York University, New York, NY, USA
\item[$^{86}$] University of Chicago, Enrico Fermi Institute, Chicago, IL, USA
\item[$^{87}$] University of Delaware, Department of Physics and Astronomy, Bartol Research Institute, Newark, DE, USA
\item[] -----
\item[$^{a}$] Max-Planck-Institut f\"ur Radioastronomie, Bonn, Germany
\item[$^{b}$] also at Kapteyn Institute, University of Groningen, Groningen, The Netherlands
\item[$^{c}$] School of Physics and Astronomy, University of Leeds, Leeds, United Kingdom
\item[$^{d}$] Fermi National Accelerator Laboratory, Fermilab, Batavia, IL, USA
\item[$^{e}$] Pennsylvania State University, University Park, PA, USA
\item[$^{f}$] Colorado State University, Fort Collins, CO, USA
\item[$^{g}$] Louisiana State University, Baton Rouge, LA, USA
\item[$^{h}$] now at Graduate School of Science, Osaka Metropolitan University, Osaka, Japan
\item[$^{i}$] Institut universitaire de France (IUF), France
\item[$^{j}$] now at Technische Universit\"at Dortmund and Ruhr-Universit\"at Bochum, Dortmund and Bochum, Germany
\end{description}

\section*{Acknowledgments}

\begin{sloppypar}
The successful installation, commissioning, and operation of the Pierre
Auger Observatory would not have been possible without the strong
commitment and effort from the technical and administrative staff in
Malarg\"ue. We are very grateful to the following agencies and
organizations for financial support:
\end{sloppypar}

\begin{sloppypar}
Argentina -- Comisi\'on Nacional de Energ\'\i{}a At\'omica; Agencia Nacional de
Promoci\'on Cient\'\i{}fica y Tecnol\'ogica (ANPCyT); Consejo Nacional de
Investigaciones Cient\'\i{}ficas y T\'ecnicas (CONICET); Gobierno de la
Provincia de Mendoza; Municipalidad de Malarg\"ue; NDM Holdings and Valle
Las Le\~nas; in gratitude for their continuing cooperation over land
access; Australia -- the Australian Research Council; Belgium -- Fonds
de la Recherche Scientifique (FNRS); Research Foundation Flanders (FWO),
Marie Curie Action of the European Union Grant No.~101107047; Brazil --
Conselho Nacional de Desenvolvimento Cient\'\i{}fico e Tecnol\'ogico (CNPq);
Financiadora de Estudos e Projetos (FINEP); Funda\c{c}\~ao de Amparo \`a
Pesquisa do Estado de Rio de Janeiro (FAPERJ); S\~ao Paulo Research
Foundation (FAPESP) Grants No.~2019/10151-2, No.~2010/07359-6 and
No.~1999/05404-3; Minist\'erio da Ci\^encia, Tecnologia, Inova\c{c}\~oes e
Comunica\c{c}\~oes (MCTIC); Czech Republic -- GACR 24-13049S, CAS LQ100102401,
MEYS LM2023032, CZ.02.1.01/0.0/0.0/16{\textunderscore}013/0001402,
CZ.02.1.01/0.0/0.0/18{\textunderscore}046/0016010 and
CZ.02.1.01/0.0/0.0/17{\textunderscore}049/0008422 and CZ.02.01.01/00/22{\textunderscore}008/0004632;
France -- Centre de Calcul IN2P3/CNRS; Centre National de la Recherche
Scientifique (CNRS); Conseil R\'egional Ile-de-France; D\'epartement
Physique Nucl\'eaire et Corpusculaire (PNC-IN2P3/CNRS); D\'epartement
Sciences de l'Univers (SDU-INSU/CNRS); Institut Lagrange de Paris (ILP)
Grant No.~LABEX ANR-10-LABX-63 within the Investissements d'Avenir
Programme Grant No.~ANR-11-IDEX-0004-02; Germany -- Bundesministerium
f\"ur Bildung und Forschung (BMBF); Deutsche Forschungsgemeinschaft (DFG);
Finanzministerium Baden-W\"urttemberg; Helmholtz Alliance for
Astroparticle Physics (HAP); Helmholtz-Gemeinschaft Deutscher
Forschungszentren (HGF); Ministerium f\"ur Kultur und Wissenschaft des
Landes Nordrhein-Westfalen; Ministerium f\"ur Wissenschaft, Forschung und
Kunst des Landes Baden-W\"urttemberg; Italy -- Istituto Nazionale di
Fisica Nucleare (INFN); Istituto Nazionale di Astrofisica (INAF);
Ministero dell'Universit\`a e della Ricerca (MUR); CETEMPS Center of
Excellence; Ministero degli Affari Esteri (MAE), ICSC Centro Nazionale
di Ricerca in High Performance Computing, Big Data and Quantum
Computing, funded by European Union NextGenerationEU, reference code
CN{\textunderscore}00000013; M\'exico -- Consejo Nacional de Ciencia y Tecnolog\'\i{}a
(CONACYT) No.~167733; Universidad Nacional Aut\'onoma de M\'exico (UNAM);
PAPIIT DGAPA-UNAM; The Netherlands -- Ministry of Education, Culture and
Science; Netherlands Organisation for Scientific Research (NWO); Dutch
national e-infrastructure with the support of SURF Cooperative; Poland
-- Ministry of Education and Science, grants No.~DIR/WK/2018/11 and
2022/WK/12; National Science Centre, grants No.~2016/22/M/ST9/00198,
2016/23/B/ST9/01635, 2020/39/B/ST9/01398, and 2022/45/B/ST9/02163;
Portugal -- Portuguese national funds and FEDER funds within Programa
Operacional Factores de Competitividade through Funda\c{c}\~ao para a Ci\^encia
e a Tecnologia (COMPETE); Romania -- Ministry of Research, Innovation
and Digitization, CNCS-UEFISCDI, contract no.~30N/2023 under Romanian
National Core Program LAPLAS VII, grant no.~PN 23 21 01 02 and project
number PN-III-P1-1.1-TE-2021-0924/TE57/2022, within PNCDI III; Slovenia
-- Slovenian Research Agency, grants P1-0031, P1-0385, I0-0033, N1-0111;
Spain -- Ministerio de Ciencia e Innovaci\'on/Agencia Estatal de
Investigaci\'on (PID2019-105544GB-I00, PID2022-140510NB-I00 and
RYC2019-027017-I), Xunta de Galicia (CIGUS Network of Research Centers,
Consolidaci\'on 2021 GRC GI-2033, ED431C-2021/22 and ED431F-2022/15),
Junta de Andaluc\'\i{}a (SOMM17/6104/UGR and P18-FR-4314), and the European
Union (Marie Sklodowska-Curie 101065027 and ERDF); USA -- Department of
Energy, Contracts No.~DE-AC02-07CH11359, No.~DE-FR02-04ER41300,
No.~DE-FG02-99ER41107 and No.~DE-SC0011689; National Science Foundation,
Grant No.~0450696, and NSF-2013199; The Grainger Foundation; Marie
Curie-IRSES/EPLANET; European Particle Physics Latin American Network;
and UNESCO.
\end{sloppypar}

}


\begin{thebibliography}{99}

\scriptsize{

\bibitem{dipole}
Aab, A. et al [Pierre Auger Coll.]
\href{https://www.science.org/doi/10.1126/science.aan4338}{Science, \textbf{357}, 6357 (2017)}

\bibitem{WaxBah}
Waxman, E. and Bahcall, J.
\href{https://link.aps.org/doi/10.1103/PhysRevLett.78.2292}{PRL, 78, 2292 (1997)}

\bibitem{Meszaros}
Mészáros, P. 
\href{https://iopscience.iop.org/article/10.1088/0034-4885/69/8/R01}{Rep. Prog. Phys., 69 2259 (2006)}

\bibitem{Fireball_revisited}
Hümmer, S. and Baerwald, P. and Winter, W.
\href{http://dx.doi.org/10.1103/PhysRevLett.108.231101}{PRL, 108, 1079 (2012)}

\bibitem{Models_NeuCosmA}
Hümmer, S. and Rüger, M. and Spanier, F. and Winter, W.
\href{http://dx.doi.org/10.1088/0004-637X/721/1/630}{ApJ, 721, 1538 (2010)}

\bibitem{Lia}
Lia, V.D. and Tamborra, I.
\href{https://iopscience.iop.org/article/10.1088/1475-7516/2024/10/054}{JCAP, 10, 054 (2024)}


\bibitem{MVT_2}
 Golkhou, V. Z. and Butler, N. R. and Littlejohns, O. M.
\href{https://dx.doi.org/10.1088/0004-637X/811/2/9}{Am. Astr. Soc., 811, 93 (2010)}

\bibitem{Ghirlanda}
{Nava}, L. and {Sironi}, L. and {Ghisellini}, G. and {Celotti}, A. and {Ghirlanda}, G.
\href{https://ui.adsabs.harvard.edu/abs/2018A&A...609A.112G}{A\&A., 609, A112 (2018)}


\bibitem{EpAlpha}
Virgili, F. J. and Qin, Y. and Zhang, B. and Liang, E.
\href{https://doi.org/10.1111/j.1365-2966.2012.21411.x}{MNRAS, 424, 2821 (2012)}

\bibitem{Pointlike}
 Aab, A. et al [Pierre Auger Coll.]
\href{https://dx.doi.org/10.1088/1475-7516/2019/11/004}{JCAP, 11, 004 (2019)}

\bibitem{Antares_Zegarelli}
Zegarelli, A. and Celli, S. et al [ANTARES Coll.]
\href{https://doi.org/10.1093/mnras/staa3503}{MNRAS, 500, 5614 (2021)}

\bibitem{IceCube}
Aartsen, M.G. et al [IceCube Coll.]
\href{https://doi.org/10.3847/1538-4357/aa7569}{ApJ, 843 112 (2017)}

}


\end{thebibliography}
\end{document}